\documentclass[aps,pra,footinbib,groupedaddress,preprint]{revtex4}

\usepackage{graphicx}
\usepackage{amsmath, epsfig}
\usepackage{amsfonts}
\usepackage{amssymb}
\usepackage{epstopdf}
\usepackage{setspace}

\newcommand{\comment}[1]{}

\newcommand{\footremember}[2]{%
   \footnote{#2}
    \newcounter{#1}
    \setcounter{#1}{\value{footnote}}%
}

\begin{document}
\title{The Spectrum of Quasistable States in a Strong Microwave Field}

\author{A. Arakelyan$^{1}$,
T. Topcu$^{2,3}$,
 F. Robicheaux$^{2}$\footremember{*}{Current address: Department of Physics, Purdue University, West Lafayette, Indiana 47907, USA}, and T.F. Gallagher$^{1}$}
\affiliation{$^1$Department of Physics, University of Virginia, Charlottesville, Virginia 22904-0714, USA \\
$^{2}$Department of Physics, Auburn University, Alabama 36849-5311, USA\\
$^{3}$Department of Physics, University of Nevada, Reno, NV 89557, USA}

\date{\today}

\begin{abstract}
When atoms are exposed to intense laser or microwave pulses $\sim$ 10\% of the atoms are found in Rydberg states subsequent to the pulse, even if it is far more intense than required for static field ionization. The optical spectra of the surviving Li atoms in the presence of a 38 GHz microwave field suggest how atoms survive an intense pulse. The spectra exhibit a periodic train of peaks 38 GHz apart. One peak is just below the limit, and with a 90 V/cm field amplitude the train extends from 300 GHz above the limit to 3000 GHz below it. The spectra and quantum mechanical calculations imply that the atoms survive in quasi stable states in which the Rydberg electron is in a weakly bound orbit infrequently returning to the ionic core during the intense pulse.
\end{abstract}

\maketitle

\section{Introduction}

When ground state atoms are ionized by an intense laser pulse, Rydberg atoms can be produced as a byproduct \cite{jones2}. Even when multiple ionization of the atom occurs, it can be accompanied by the production of ionic Rydberg states converging to higher ionization limits \cite{PhysRevLett.93.023001}. While the notion of producing weakly bound Rydberg states with an intense laser pulse might seem surprising, it can be understood in the following way. The intense laser pulse ejects the electron from the ion core either in a single field cycle, through tunnel ionization, or over many cycles, via multiphoton ionization (MPI). In either case, the electron departs from the core with a range of energies into the long range tail of the coulomb potential, where the electron is nearly free, and there is no cycle average energy exchange with the laser field.  Above the ionization limit the range of energies appears explicitly in the form of above threshold ionization (ATI) \cite{PhysRevLett.42.1127}, but the energy range extends below the limit as well. An electron with an energy below the limit is reflected by the coulomb potential and returns to the core in one Kepler period \cite{jones2, jones, stapelfeldt}. If the laser pulse is over before the return, the electron obviously remains bound. If the laser pulse is not over, rescattering of the electron from the core can lead to superponderomotive ATI, recombination and high harmonic generation (HHG), non sequential double ionization (NSDI), and nearly elastic scattering. Nearly elastic scattering can result in a very low energy free electron, as observed recently \cite{blaga,PhysRevLett.103.093001,PhysRevLett.110.013001}, or as a bound electron \cite{PhysRevLett.42.1127,McPherson:87,PhysRevLett.69.2642,PhysRevLett.101.233001,eichmann}. A striking example of the last process is provided by recent experiments with He in which 10\% of the surviving atoms are attributed to states of $n<7$ with Kepler orbital periods less than the 30 fs duration of the 1GV/cm laser pulse \cite{eichmann}.

How atoms survive an intense field was first addressed in calculations of high frequency stabilization \cite{PhysRevLett.21.838, PhysRevLett.61.939}. More recent calculations to elucidate the source of superponderomotive electrons in ATI are, however, more relevant to this problem~\cite{PhysRevLett.83.3158}. Superponderomotive electrons are those with energies in excess of the prediction of the Simpleman's Model \cite{PhysRevLett.61.2304,PhysRevLett.62.1259,vlvdh}. They are formed at the intensities at which the AC Stark shift from the laser field brings states just below the limit into multiphoton resonance with the ground state. At these intensities quasi stable states are produced in which the atom survives multiple recollisions with the core, with some loss of population on each recollision.

Microwave ionization of Rydberg atoms exhibits many similarities to MPI of ground state atoms. In particular, recent microwave ionization experiments have shown that 10\% of the atoms remain bound in high lying states at fields orders of magnitude beyond those required for static field ionization. Furthermore, electrons in the high lying states remain bound even if they return to the ionic core during the microwave pulse, essentially the same result as observed in the laser experiments \cite{PhysRevLett.83.3158}. While quasi stable high lying states are formed in both the laser and microwave experiments, an attractive feature of the microwave case is that it is possible to probe them spectrally. Here we report the optical excitation spectra of Rydberg atoms in a strong microwave field. Our spectra and calculations demonstrate that the surviving atoms result from excitation to microwave dressed states just below the limit. Specifically the spectra consist of series of peaks separated by the microwave frequency. These observations imply that the atoms are in quasi stable states in which the stability derives largely from the electron's spending most of its time far from the ionic core. Nonetheless, the atom survives the electron's revisiting the core. In the sections which follow we describe our experimental approach, present our experimental observations, and show the results of one dimensional quantum calculations of the spectra.

 \section{Experimental approach}

\begin{figure}[eps,pra]
\begin{center}
\centerline{\scalebox{0.3}{\includegraphics{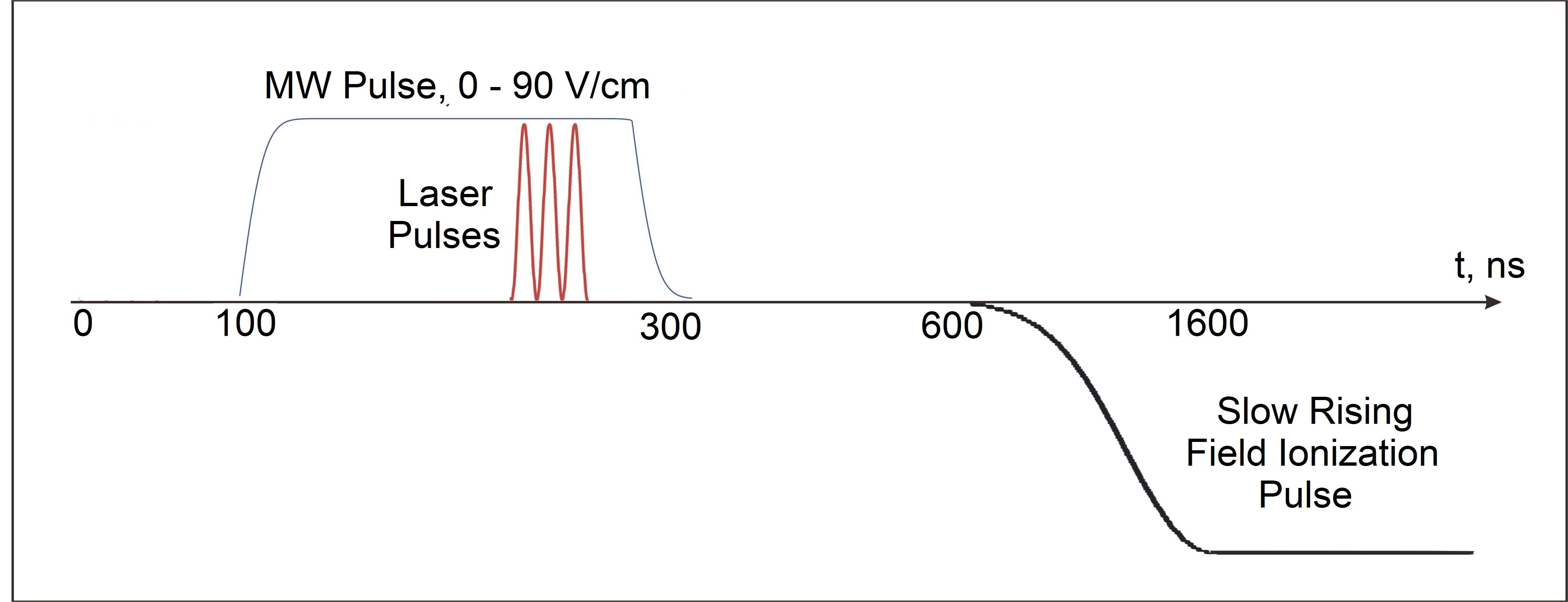}}}
\caption{(Color online).
Experimental timing diagram for the investigation of the microwave recombination in Li by a 38.3 GHz microwave (MW) field}
\label{timing}
\end{center}
\end{figure}

This section provides only a brief outline of the experimental apparatus used in the current experiment since a detailed description can be found elsewhere ~\cite{arakelyan}. A thermal beam of Li atoms in its ground state passes through the antinode at the center of a 38.34-GHz Fabry-Perot cavity where the atoms are excited to the vicinity of the ionization limit by three 20-ns dye laser pulses via the $2s\rightarrow2p\rightarrow3s\rightarrow np$ transitions at wavelengths of 670, 813, and 615 nm, respectively. The experiment is run at a 500 Hz repetition rate. The laser excitation occurs in the presence of a microwave field, as shown in the timing diagram in Fig. \ref{timing}, so the oscillator strength to the $np$ states becomes nearly continuously distributed in energy. The microwave pulse starts before the laser excitation and lasts for a minimum of 20-ns after it. Unless stated otherwise, the microwave field is on for 30 ns after the pulsed laser excitation. The rise and fall times of the pulse are approximately 20 ns each. After the end of the microwave pulse, typically 300 ns later, atoms are field
ionized by a field ionization pulse. The electrons ejected by the ionization field are detected by a dual micro-channel plate detector (MCP). The time resolved MCP signal tells us how many atoms survived the microwave pulse in bound states, as well as their final state distribution. The MCP signal is captured by a gated integrator or oscilloscope and recorded for later analysis. The microwave system generates a 38.34-GHz pulse with a variable width and 0- to 90 V/cm amplitude which is determined an uncertainty of 15$\%$. A stray field of 8 mV/cm is present in the excitation volume, and well defined static fields can be added. A relative frequency measurement of the $3s\rightarrow np$ laser is obtained with the help of a 52.42-GHz free spectral range etalon, and an optogalvanic signal from the 16274.0212-cm$^{-1}$ 2p$^5$(2p$_{3/2}$)3s-2p$^5$(2p$_{3/2}$)2p Ne line provides an absolute calibration. The three laser beams are focused to less than 1-mm diameter spots where they intersect the atomic beam. The 615-nm beam propagates perpendicularly to the 670 and 813 nm beams, so the volume of excited atoms is $\approx$ 1 mm$^3$. The laser field and microwave field are polarized vertically.

 \section{observations}

\begin{figure}[eps,pra]
\begin{center}
\centerline{\scalebox{0.5}{\includegraphics{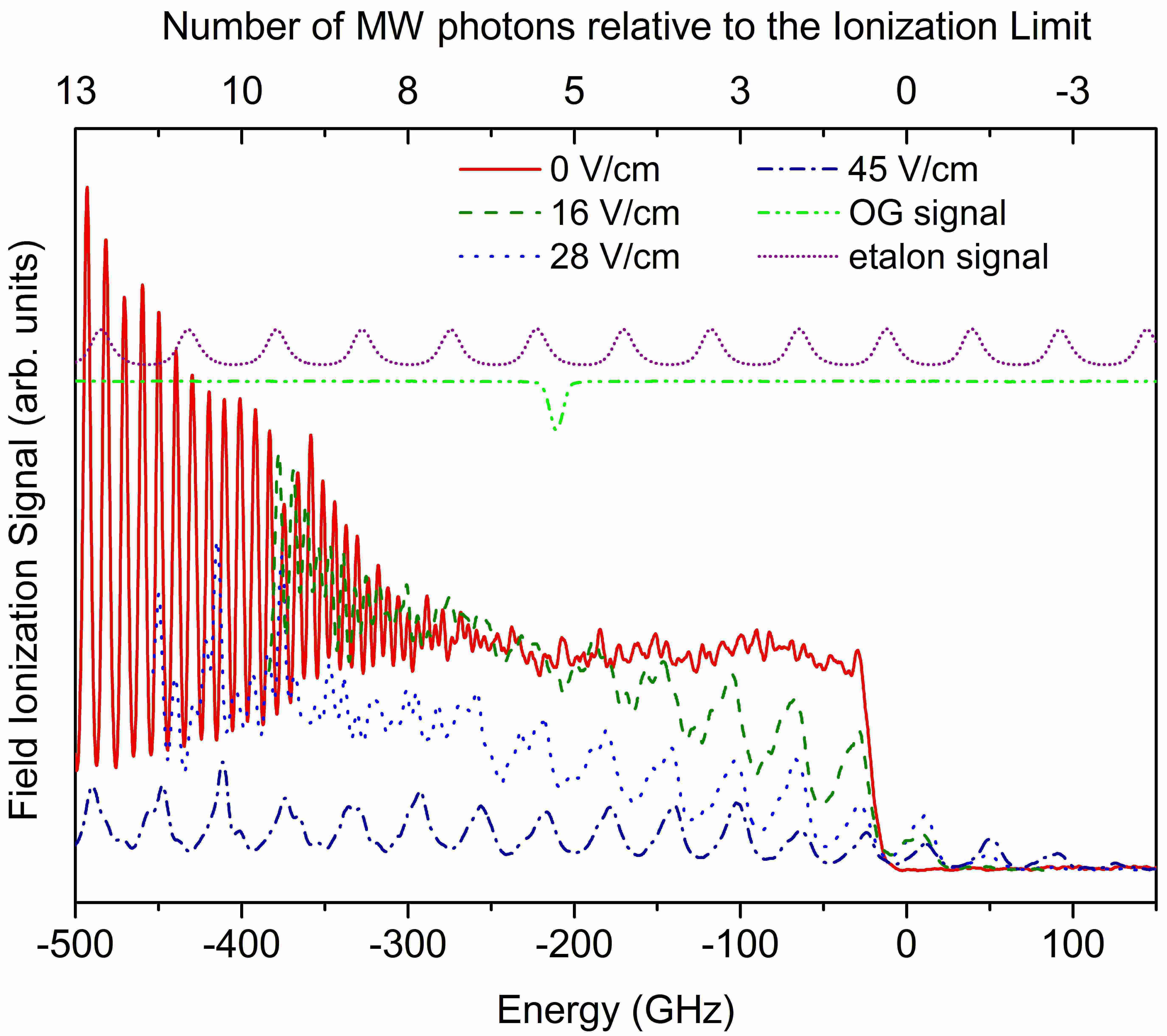}}}
\caption{(Color online).
Field ionization signal as a function of laser frequency tuning, in terms of energy relative to the ionization limit (IL), after exposure to a microwave pulse of amplitude: 0 V/cm, 16 V/cm, 36 V/cm, and 45 V/cm. The horizontal axis is calibrated by the etalon and optogalvanic signals.}
\label{zerofileldlowpower}
\end{center}
\end{figure}

In Fig. \ref{zerofileldlowpower} we show the spectra obtained by scanning the frequency of the third laser in the presence of microwave fields from 0 to 45 V/cm. The gate of the integrator is wide enough to collect the field ionization signals of all states of $n>43$, that is of energies greater than $-1500$ GHz. The zero of energy of the horizontal scale is the ionization limit. In zero microwave field we observe resolved bound Rydberg states at energies lower than $-300$ GHz and a flat signal at higher energies, where the spacing between adjacent levels is smaller than the 8-GHz linewidth of our laser, which occurs at $n\approx100$. At an energy of $-45$ GHz, the signal starts falling to zero because photoelectrons are not detected when the laser is tuned over the ionization limit. Energies are measured relative to the zero static field limit. Due to the finite
linewidth of the laser, the signal drops to zero with a finite slope, and we denote the middle of that slope as the ionization limit. The limit is depressed due to the stray static field in the interaction region mentioned above. We estimate the stray field using $\Delta W = 2\sqrt{E}$. Typically, the data are taken when the depressed ionization limit is at $-18$ GHz, which corresponds to a stray field of 8 mV/cm. As shown by Fig. \ref{zerofileldlowpower}, the presence of the microwave field leads to a noticeable modulation of the spectrum at the microwave frequency. There are several points to note regarding the modulation. In the 16 V/cm spectrum of Fig. \ref{zerofileldlowpower} the high frequency side of the peak at the limit coincides with the drop in signal at the limit in the zero (microwave) field spectrum of Fig. \ref{zerofileldlowpower}. In addition, the peaks in the signal near the limit are asymmetric, but the asymmetry disappears at higher microwave fields. As the microwave field is raised the
regular structure persists further both above and below the ionization limit, as observed previously~\cite{gurian,shuman,Overstreet:106}. This asymmetry is expected if the ionization occurs through photon absorption in the perturbative limit (low power) since in this case the photoabsorption rate scales as $1/n^3$.
This results in a higher photoionization rate for an electron bound by 95\% of the photon energy than for one
bound by 5\% of the photon energy. In addition, the entire set of peaks moves to higher energy due to the ponderomotive energy shift. For the highest microwave field we have used, 90 V/cm, we observe a ponderomotive shift of 15 GHz, confirming our field calibration.

\begin{figure}[eps,pra]
\begin{center}
\centerline{\scalebox{0.3}{\includegraphics{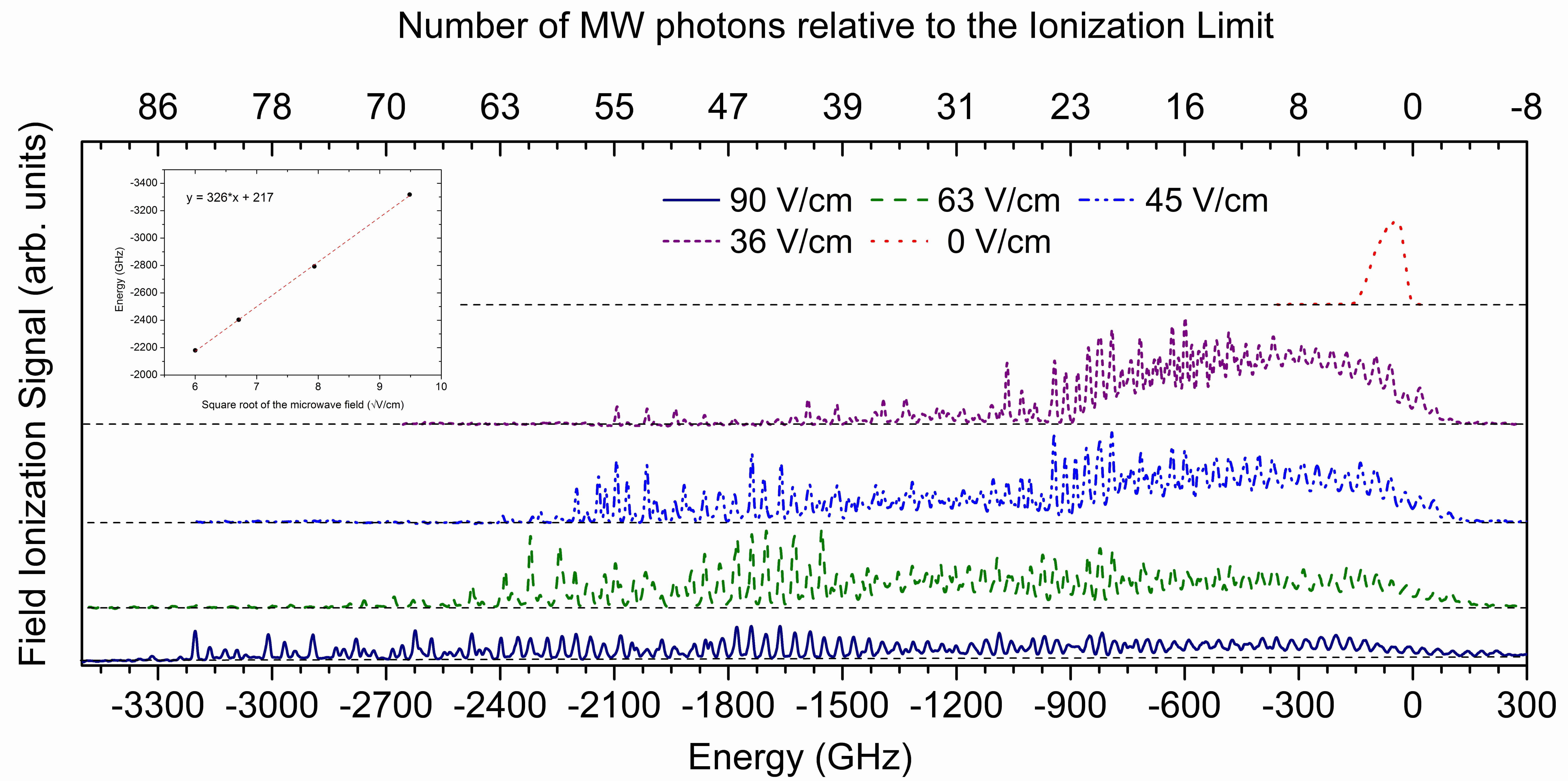}}}
\caption{(Color online).
Field ionization signal as a function of laser frequency tuning, after exposure to a microwave pulse of amplitude: 0 V/cm, 36 V/cm, 45 V/cm, 63 V/cm, and 90 V/cm. A 50-ns integration gate is used to detect only atoms bound within 75 GHz of the IL. The zero field spectrum is scaled by a factor of 3 to fit the graph. Horizontal dashed lines show the zeroes of the corresponding spectra. The inset shows the linear behavior of the onset of the microwave structure as a function of square root of the MW field with a dependence of $W = 4\sqrt{E}$. Apparently, all bound states in the microwave structure at any MW field are extremely high lying states trapped within two MW photons of the IL ($n>215$).  }
\label{slow}
\end{center}
\end{figure}

Inspection of the time resolved field ionization signal obtained in the 45-V/cm trace of Fig. \ref{zerofileldlowpower} shows that all the signal comes from atoms in very highly excited states. Accordingly we narrowed the gate of the integrator to 50 ns to detect only atoms bound by less than 75 GHz and recorded the spectra shown in Fig. \ref{slow}. When there is no microwave field, we only detect a signal when the laser is tuned within 75 GHz of the depressed ionization limit at $-18$ GHz.  As the microwave field amplitude is increased the spectrum extends progressively further above the limit, and more interesting, to very deeply bound states. In the inset we plot the binding energy at which each spectrum appears vs the square root of the microwave field, which gives a linear dependence.  Converting from the laboratory units of the inset to atomic units gives $W=4\surd E$ almost exactly twice the binding energy for classical field ionization. Whether or not the $W=4\surd E$ dependence shown in the inset of Fig. \ref{slow} is frequency dependent is an open question.

\begin{figure}[eps,pra]
\begin{center}
\centerline{\scalebox{0.5}{\includegraphics{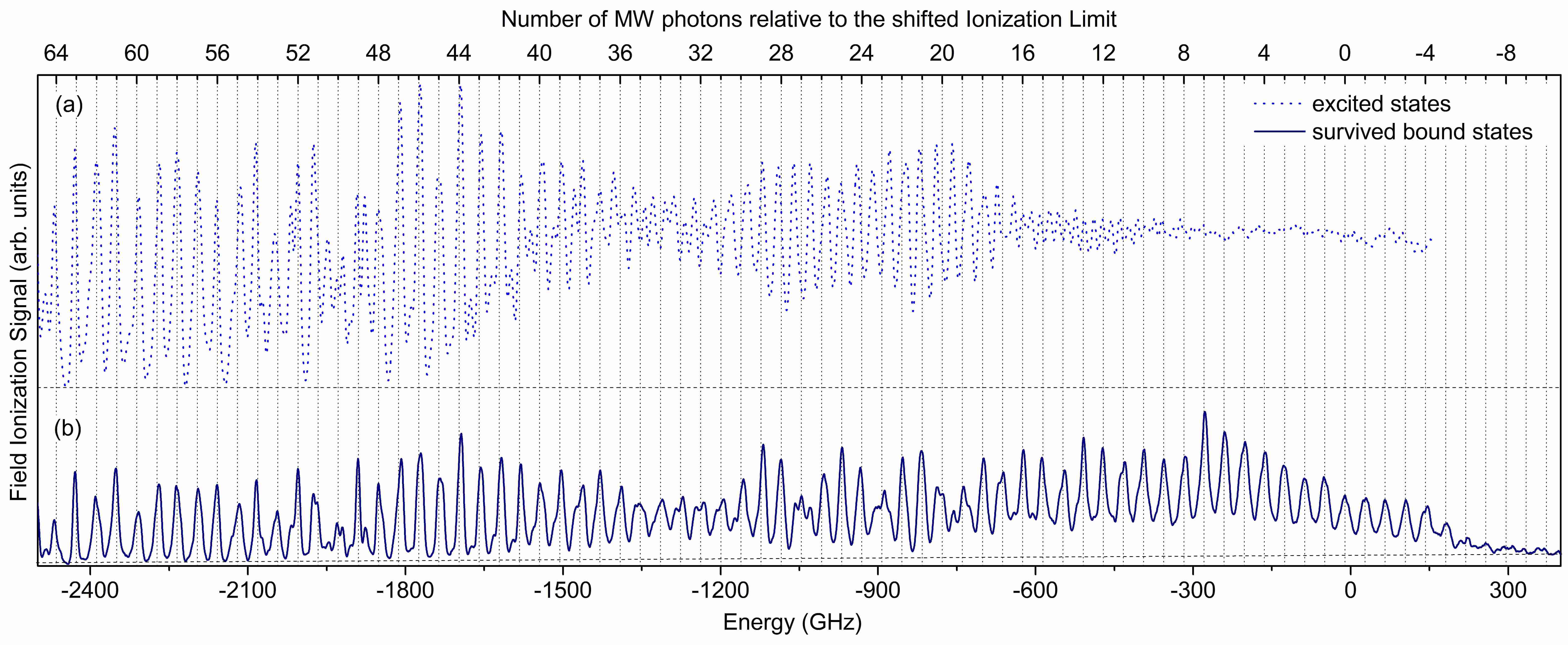}}}
\caption{(Color online).
Field ionization signals as a function of laser frequency tuning for excitation in the presence of a microwave pulse of amplitude of 90 V/cm. (a) The total photoabsorbtion spectrum when detecting all energies above $n=43$ ( dotted line).  (b) Spectrum obtained when detecting only the high lying states 9 solid line). The total photoabsorbtion spectrum is scaled by a factor of eight to fit the graph. Horizontal dashed lines show the zeroes of the corresponding spectra. The total photoabsorption spectrum also presents the 38 GHz structure at energies below $-1500$ GHz since only atoms transferred to higher lying states are detected.  }
\label{alle}
\end{center}
\end{figure}

The regularity of the structure at the microwave frequency extends almost as far as the spectra shown in Fig. \ref{slow}.  In Fig. \ref{alle}b we show an expanded view of the 90-V/cm spectrum of Fig. \ref{slow}, as well as a set of vertical dotted lines spaced by the microwave frequency to show the regularity of the microwave induced structure down to an energy of $-2400$ GHz. In Fig. \ref{alle}a we show the spectrum recorded in the presence of a 90-V/cm microwave field when the atoms are ionized immediately after the microwave pulse by a field pulse which ionizes atoms bound by up to 1500 GHz (Detecting ions yields the same result.). At energies below $-1500$ GHz not all the initially excited atoms are detected, only ones which have been transferred to within 1500 GHz of the ionization limit. Thus the spectrum has the same modulation as the spectrum of Fig. \ref{alle}b.  At energies above $-1500$ GHz there is in general no obvious connection between the spectra of Fig. \ref{alle}a and \ref{alle}b.  An
exception is the region near $-900$ GHz where there is a modulation in the spectrum of Fig. \ref{alle}a at a slightly lower frequency than the microwave frequency.  The microwave frequency matches the Kepler, $\Delta n=1$, frequency at $-1050$ GHz, which is probably the origin of the structure in Fig. \ref{alle}a. In any event, the deep modulation of the total number of atoms excited, shown in Fig. \ref{alle}a, introduces a beat note into the spectrum of Fig. \ref{alle}b.  Finally, the spectrum of Fig. \ref{alle}b extends from the regime in which the microwave frequency is much larger than the $\Delta n=1$ frequency, at the limit, to the regime in which it is smaller than the $\Delta n=1$ frequency. For example, the Li 37p and 38p states, bound by 2409 and 2284 GHz, respectively, are separated by three times the microwave frequency.

\begin{figure}[eps,pra]
\begin{center}
\centerline{\scalebox{0.50}{\includegraphics{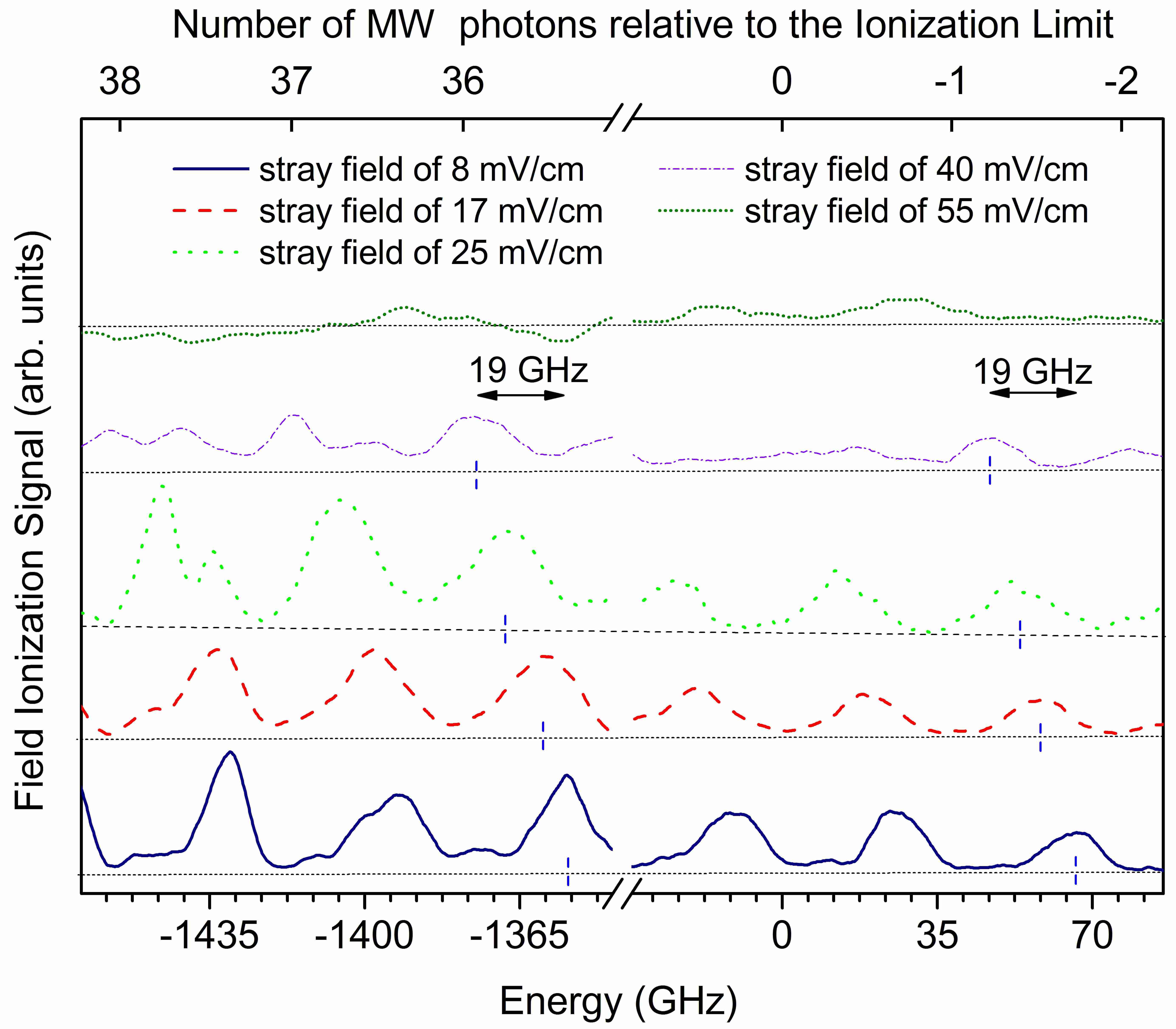}}}
\caption{(Color online).
Field ionization signal as a function of laser frequency tuning after exposure to a microwave pulse of amplitude of 90 V/cm for different static fields in the interaction region: 8 mV/cm, 17 mV/cm, 25 mV/cm, 40 mV/cm, and 55 mV/cm. A break is shown on the horizontal axis for convenience. Horizontal dashed lines show the baselines of the corresponding spectra. Vertical blue lines intersect the horizontal baselines at points to which the ionization limit is depressed due to $W=2\surd E$ for the corresponding static field $E$. It is apparent that peaks in the microwave structure also shift by $W$, irrespective of the laser frequency. Increasing the static field destroys the microwave structure, and the 55-mV/cm spectrum shows only noise,with no bound population detected. }
\label{strayfields}
\end{center}
\end{figure}

Several observations suggest that the spectra are tied to the ionization limit. First, the spectra of Fig. \ref{slow} were recorded by detecting very high lying states. Second, in Figs. \ref{zerofileldlowpower} and \ref{slow} the extent of the spectra above and below the limit increases with the microwave field amplitude. Finally, the 16-V/cm spectrum of Fig. \ref{zerofileldlowpower} appears to have a peak at the ionization limit.  As an explicit test of this notion we have recorded spectra in the presence of small static fields which depress the limit by different amounts. In Fig. \ref{strayfields} we show segments of spectra recorded in static fields of up to 55 mV/cm. In particular, we show segments of the spectra at the limit and at the energy $-1400$ GHz. At the limit the spectra shift to lower energy as the depressed limit shifts, and precisely the same shift is observed at $-1400$ GHz. In short, the entire spectrum is tied to the ionization limit. The other important aspect of Fig. \ref{strayfields}
is that the size of the peaks decreases with increasing static field, and the peaks essentially disappear at 55 mV/cm.

The dependence on the static field suggests that in the presence of the microwave field the stable electron orbits are large. In particular, in a static field the coulomb potential is no longer infinite in range but has a finite range, given by $z=1/\surd E$. A field of 50 mV/cm corresponds to $z=3\times 10^5$ a$_0$ and a Kepler orbital period of 3 ns, and it seems possible that stable orbits require a larger volume or longer orbit time than allowed by this field.

\begin{figure}[eps,pra]
\begin{center}
\centerline{\scalebox{0.33}{\includegraphics{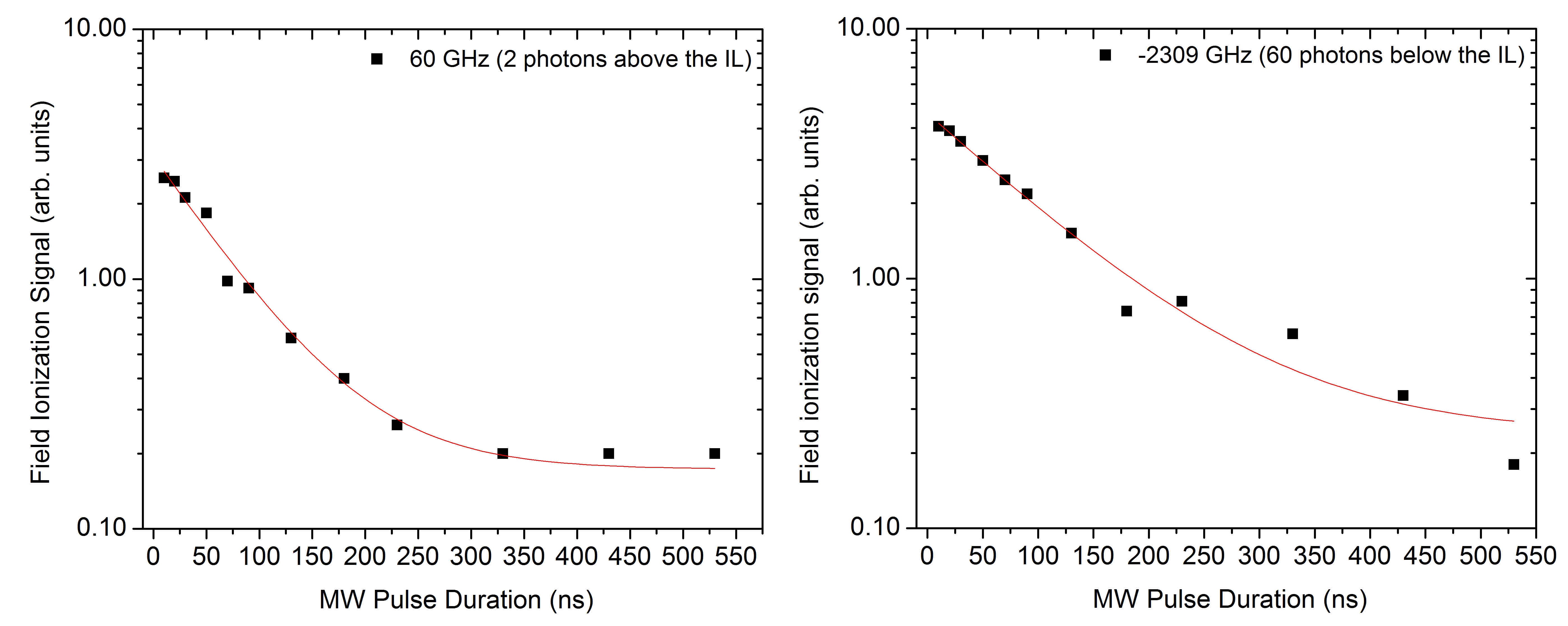}}}
\caption{(Color online).
Field ionization signal as a function of microwave pulse duration after the laser excitation. The microwave pulse amplitude is 90 V/cm, and the laser frequency is fixed and tuned to (a) two mw photons above the ionization limit and (b) 60 photons below it. Only the very high lying states are collected with a 50-ns gate. The solid line represents a fit of the form $Ae^{-x/t}+c$. According to the fit, the lifetimes of decaying exponentials are 70 and 110 ns, respectively. The lifetimes  exhibit an initial rapid decay followed by a very slowly decaying tail.  }
\label{lifetime}
\end{center}
\end{figure}

We  have measured the lifetimes in the microwave field by varying the exposure time of the atoms to the microwave field subsequent to laser excitation. Specifically, with the third laser tuned to a peak in the spectrum of Fig. \ref{alle}b we increased the microwave pulse width in small increments, with a fixed delay between the end of the microwave pulse and the FIP, and recorded the field ionization signal. In general we observed a two component decay, with a rapid initial decay followed by a slowly decaying tail, as shown in Fig. \ref{lifetime} for two laser tunings. The rapid initial decay time is $\sim$ 70 ns at the limit and increases to 110 ns for the most deeply bound peak at $-2309$ GHz. The long tail has a decay time of $\sim$ 1 $\mu$s, irrespective of laser tuning. Although the decays shown in Fig. \ref{lifetime} were obtained by tuning the laser to the peaks of Fig. \ref{alle}b, tuning the laser midway between the peaks leads to a similar result. There is no obvious dependence of the lifetime on
the microwave field amplitude if the modulation in the excitation spectrum is visible. Although a two component decay suggests the possibility that peaks such as those shown in Fig. \ref{alle}b might become narrower with longer exposure to the microwave field, such is not the case. The peaks in the spectrum have the same width with 30 ns and 5400 ns of microwave pulse subsequent to laser excitation.

\section{One-dimensional quantum simulations}

To provide further insight into the spectra we have carried out
one dimensional quantum calculations. The initial calculations were
done for a microwave frequency of 38 GHz. While these calculations
gave results in accord with the experimental results, they were
barely converged, which precluded our probing subtle features with
confidence. To circumvent this obstacle we scaled the problem. We have used $n$ states a factor of ten lower, $n\approx20$ instead of $n\approx200$, a microwave frequency of 38 THz instead of 38 GHz,  and a field of $9\times 10^5$ V/cm instead of 90 V/cm. With this scaling we
are able to obtain well converged results. We numerically solve the time-dependent Schr\"odinger equation using a Green's function method with
wave packets launched in a continuous wave (cw) microwave field. The time-dependent
Schr\"odinger equation to be solved is
\begin{equation}\label{scheq}
\text{i}\frac{\partial\psi(x,t)}{\partial t} - (H-E)\psi(x,t)=S(x,t)
\end{equation}
where $H$ is Hamiltonian, and $E$ is the energy at which the wavepacket is launched, and the wave function $\psi(x,t)$
is initially zero everywhere. We use a Numerov scheme for our nonlinear spatial mesh to evolve
Eq. (\ref{scheq}) in time; as described in detail in Ref.~\cite{robicheaux}.
The duration of the
source term $S(x,t)$ is 3 microwave periods (giving $\sim$8 THz at FWHM), and it has a Gaussian envelope in
time. The wave packets are launched with energies ranging from -140 THz to 60 THz relative to the
ionization threshold. We use a soft-core potential given by $V(x)=-1/\sqrt{x^2+a^2}$ with $a=1$ a.u. Since $V\rightarrow -1/|x|$ at large $|x|$, the potential has an infinite number of bound states. However, since the wavefunction exists for $x<0$, there are twice as many states as in the H atom. For example, in this potential an atom with the H $n=10$ binding energy has $n=20$. The depth of the potential $a$ has no direct consequence
for the physical process of interest, the trapping of population occurs in
weakly bound states that have a spatial extent much larger than $a$.

\begin{figure*}[!ht]
	\begin{center}
		\resizebox{120mm}{!}{\includegraphics{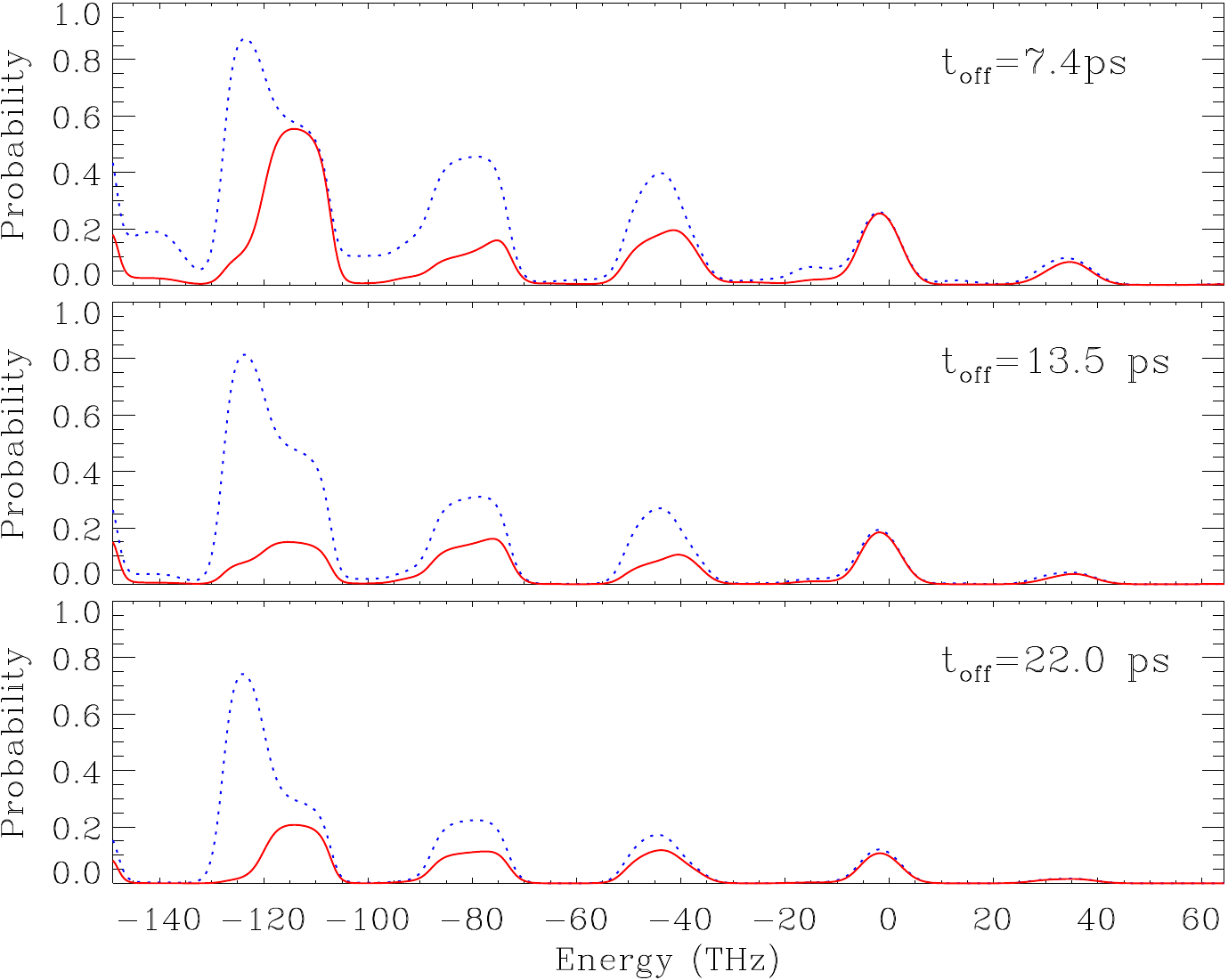}}
	\end{center}
	\caption{(Color online). Calculated spectra for a $9\times 10^{5}$ V/cm 38 THz
field from scaled one-dimensional quantum simulations at three different turn-off times for
	the microwave field. We smoothly turn off the microwaves to mimic the experimental
	conditions and wait until the field strength drops below $10^{-8}$ V/cm before we
	compute the surviving norm. The dotted blue curves show the spectra obtained when all bound states are detected, and the solid red curves show the spectra obtained when only bound atoms within 38 THz of the limit are detected.
	}
	\label{norm:scaled}
\end{figure*}

The calculations lead to bound state fractions similar to those seen in the experiment. In Fig. \ref{norm:scaled} we show the calculated spectra of bound atoms found after excitation in a 38 THz field of $9\times10^5$ V/cm. Specifically, we show the spectra for all remaining bound atoms (dotted blue curve) and those within one 38 THz photon of the ionization limit (solid red curve). As shown by Fig. \ref{norm:scaled}, the spectra are similar to the experimental spectra; peaks are found at the ionization limit modulo the microwave frequency. Fig. \ref{norm:scaled} also provides information about the lifetime in the microwave field. When the atoms are exposed to longer 38 THz pulses the number of bound atoms decreases. An important point is that as the microwave pulse is lengthened the total number of bound atoms decreases more rapidly than the number of atoms within one microwave photon of the limit. When excitation occurs at the limit the high lying states have a lifetime of approximately 18 ps. In view of the three order of magnitude time scaling, this lifetime is consistent with the experimental lifetimes shown in Fig. \ref{lifetime}. In long microwave pulses the bulk of the surviving atoms are in states whose energies are within one microwave photon of the limit, as observed experimentally.

\begin{figure*}[!ht]
	\begin{center}
		\resizebox{110mm}{!}{\includegraphics{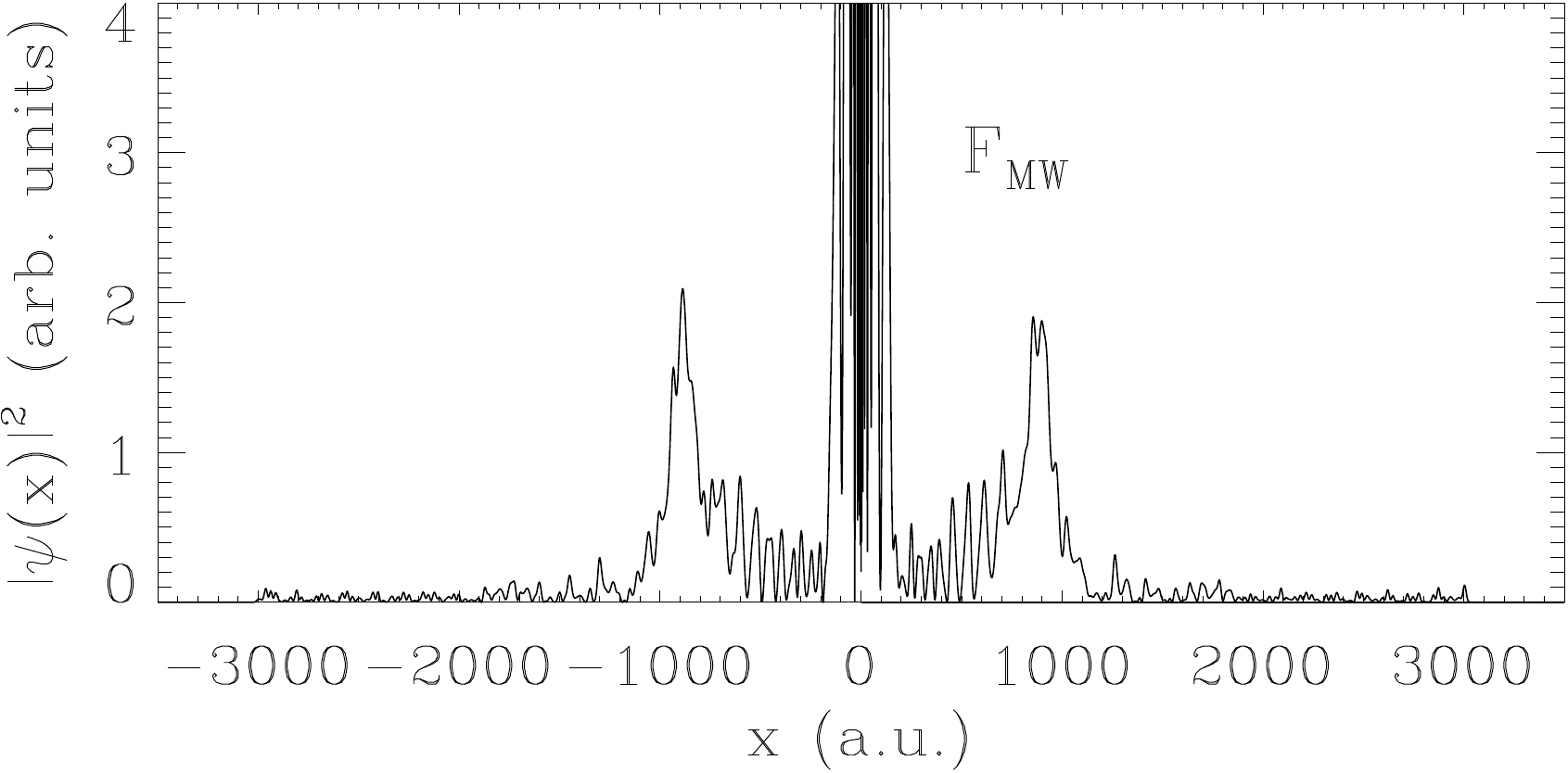}}
	\end{center}
	\caption{(Color online). The squared wavefunction 256 microwave cycles after the excitation pulse.
	There is subtantial amplitude at small $x$ and again at $x\sim\pm 900$ a.u.
	The peaks at $\pm 900$ a.u. are persistent, whereas the small $x$ part of the wavefunction decreases for sufficiently
	long microwave exposure. The peaks at $x=\pm900$ a.u. are located at the classical turning points of $n=21$.
	}
	\label{fig:wfunc}
\end{figure*}

The similarity of the calculated spectra to those observed experimentally gives us confidence that this simple model captures the essential physics of this system. Thus, we can use the wavefunction to help understand
the unexpected stability of these states. Fig.~\ref{fig:wfunc} shows the spatial probability distribution for the wavefunction within 3000~a$_0$ of the nucleus 256 microwave cycles, 6.7 ps, after the excitation pulse. There is a large probability for the electron to be at
$|x|\sim 900$~a$_0$, which corresponds to a few zero field states of $n\sim 21$.
A state with an outer turning point of $\sim 900$~a$_0$ is bound by 3.6 THz, a binding small compared to 38 GHz. The $n=21$ Kepler period is $2.8$~ps, a factor of three shorter than the exposure time to the microwaves. As in the experiment, the electron remains bound even though it returns to the core during the microwave pulse. There is also substantial population at small $x$, a feature that is observed in the experiment only at small microwave fields and only for a range of initial binding energies where orbital frequency is larger than the microwave frequency. This population corresponds to $n\simeq5$, where the Kepler frequency is comparable to the microwave frequency. The population at small $|x|$ decays more rapidly than that at large $|x|$ and may be an intermediate state in their ionization.

We have also performed one- and three-dimensional Classical Trajectory
Monte Carlo (CTMC) simulations to further investigate the origin of this effect. However, in none of our classical simulations, have we observed any surviving atoms, suggesting that the survival
is entirely quantum mechanical in origin. Moreover, $\sim$51\% of all trajectories
are within 38 GHz {\it above} the threshold. The classical piling up of the atoms within
a single photon {\it above} threshold could mean piling up within a single photon
{\it below} the threshold quantum mechanically. Unlike in classical mechanics, in quantum mechanics
the electron has to absorb energy in discrete units (photons) and has to pass near the
nucleus to absorb a photon. If the electron is in a state with a classical orbital period
comparable or larger than the microwaves duration, then quantum mechanically it will
not have a chance to absorb the final necessary photon to escape. This, however,
classically poses no problem for the escape of the electron, because when the electron
acquires enough energy which puts it within less than a photon from the threshold, it
only needs a fraction of a photon's energy to go above above the threshold, which it can
absorb continuously without need for passing close to the nucleus. Indeed, our quantum
simulations have revealed this to be the mechanism for the survival of the atoms in the
microwave field.

\section{conclusion}

These measurements and calculations show that the surprising observation of highly excited bound states after the exposure of atoms to a strong laser or microwave pulse is due to the excitation of weakly bound states just below the ionization limit. Such excitation is possible at any binding energy, as long as it is modulo a microwave photon from the ionization limit.  Here we have explicitly shown the importance of resonant excitation by using laser excitation in the presence of a cw microwave field. In the usual case of a pulse of microwaves or laser light, the Stark shift due to the changing intensity during the pulse leads to resonant excitation of the high lying states, as shown by the calculations of Muller \cite{PhysRevLett.83.3158}.

\begin{acknowledgments}
It is a pleasure to acknowledge stimulating conversations with R. R. Jones and the support of the National Science Foundation under grant PHY-1206183. T. T. and F. R. are supported by the
Office of Basic Energy Sciences, U.S. Department of Energy. Part of the computational work was carried out at the National Energy Research Scientific Computing (NERSC) Center in Oakland, CA and on the Kraken XT5 facility at the National Institute for Computational Science (NICS) in Knoxville,
TN, as part of the XSEDE program.
\end{acknowledgments}

\bibliography{papers,theory_refs}

\pagebreak

\onecolumngrid

\end{document}